\documentclass[showpacs, twocolumn]{revtex4-1}
\usepackage{graphicx}
\usepackage{amsmath}
\usepackage{appendix}
\usepackage{array}
\usepackage{multirow}

\begin{document}

\title{ A charged Coulomb Bose gas with dipole-dipole interactions}

\author{Abdel\^{a}ali Boudjem\^{a}a}

\affiliation{Department of Physics, Faculty of Exact Sciences and Informatics, Hassiba Benbouali University of Chlef, P.O. Box 78, 02000, Ouled-Fares, Chlef, Algeria.}
\email {a.boudjemaa@univ-chlef.dz}

%\date{\today}

\begin{abstract}

We systematically study the properties of a charged Coulomb Bose gas with dipole-dipole interactions in the weak coupling limit
at both zero and finite temperatures using the Hartree-Fock-Bogoliubov approach.
We numerically analyze the collective excitations, the condensate fraction, the depletion, the chemical potential, and the static structure factor.  
Moreover, we compare our new findings with those of nondipolar charged Coulomb Bose gas.
Our results reveal that the complex interplay of  Coulomb and dipole-dipole interactions may modify the stability,  the thermodynamics and the coherence of the system.

\end{abstract}

\maketitle

%\section{Introduction}

The last decades have witnessed a remarkable surge of interest in charged bosons.
A charged Bose gas (CBG) in which particles interact via Coulomb forces was stimulated by potential applications in  statistical mechanics \cite{Wu}, 
physics of high-temperature superconductivity \cite{Alex}, Meissner-Ochsenfeld effect \cite{Shf1,Shf2}, collective excitations \cite{Bohm,Bon},
nuclear reactions in dense plasmas and astrophysics \cite{Nin,Ginz,Hans,Schr,Mul}, Wigner crystallization \cite{Wig,Cer,Pel,Drum} and so on.

The properties of CBG at zero temperature have been extensively studied using different approaches.
In 1961, Foldy \cite{Foldy} calculated the ground-state energy and the elementary excitations spectrum of CBG employing the Bogoliubov theory \cite{Bog}, valid only
at very low temperatures and in the high-density (weak coupling) regime, $r_s \ll 1$ \cite{Foldy}.
The coupling strength is characterized by the dimensionless gas parameter $r_s=r_0/a_B$, where $a_B$ is the Bohr radius,
and $r_0=(3/4\pi n)^{1/3}$ is the interparticle separation with $n$ being the mean density. 
Higher-order corrections to the ground-state energy were obtained in \cite{Lee,Bru,Woo} using beyond the Bogoliubov approximation. 
The subsequent studies dealt with the random phase approximation dielectric response function \cite{Sing,Vash,Hir},
quantum-to-classical mappings \cite{Perr,Sand}, collective modes and screening properties of CBG \cite{Alex1,Alex2}. 
Further investigations have been performed at finite temperature focusing on the critical temperature, elementary excitations, 
the normal and anomalous momentum distributions \cite{Hans,Fett,Bish,Hore,Strep,Dav,Dav1}.
Rigorous results for various ground-state properties of CBG  have been obtained in the frame of Quantum Monte Carlo methods (see e.g. \cite{Cer,Mor,Nor,Palo} and references therein).
Ground-state energies of the two-component CBG have been computed by Lieb {\it et al}.\cite{Lieb} using Dyson's method. 
The authors of Refs.\cite{Gu1,Gu2} have analyzed the magnetic properties of CBG within the mean-field theory.
Other aspects of CBG have been studied in \cite{Luk} and references therein.

In this Letter we investigate  the ground-state properties of CBG with short-range (contact) and dipole-dipole interactions (DDI)
using the full HFB theory.  Meissner effect in a charged Bose gas with short-range repulsion has been addressed in \cite{Shun}, where
the collective excitations due to the repulsive interaction found to complicate the situation. 
Ultracold quantum gases with DDI  have attracted tremendous interests recently due to the long-range character and the anisotropy 
(see for review \cite{Pfau,Carr,Baranov,Pupillo2012} and references therein) in contrast to the short-range interactions. 
Dipolar Bose-Einstein condensates (BECs) consist of atoms with sizeable magnetic dipole moments and have been experimentally realized
with ${}^{52}$Cr \cite{Gries}, ${}^{164}$Dy \cite{ming},  and ${}^{168}$Er \cite{erbium}. 
Most recently, Er-Dy mixture has been experimentally achieved in two-species magneto-optical trap \cite{Ilz}.
The DDI may strongly affect the excitations, the dynamics and  the thermodynamic properties of the BEC \cite{Pfau,Carr,Baranov,Pupillo2012}.
In addition, they lead to the emergence of novel quantum phases such as supersolid and droplet states (see for review \cite {Luo,Pfau2,Guo} and references therein).
Therefore, it is instructive to discuss the role played by the competition between the Coulomb, contact and dipolar interactions in the CBG.

Within the Hartree-Fock-Bogoliubov (HFB) theory we write down the generalized nonlocal Gross-Piteavskii and calculate the Bogoliubov excitations energy. 
We show that this latter presents a plasmon gap spectrum at long wavelengths (low momenta) regime due to the Coulomb interactions. 
The presence of DDI lead to shift the frequency of such a gap.
We provide useful analytic expressions for the normal and anomalous fluctuations, the equation of state (EoS),  and the static structure factor.
%The role of the normal and anomalous fluctuations in a nondipolar CBF at finite temperature has been examined by Davoudi {\it et al}.\cite{Dav} employing the HFB method. 
The obtained expressions (notably those of the anomalous density and the EoS) suffer from both infrared  and ultraviolet divergences. 
The former originates from the Coulomb interaction while the latter is  caused by the use of a contact interaction.
In the absence of  contact interactions and DDI, exact cancellation of infrared-divergent terms in the HFB shift of the single-particle excitation energy has been demonstrated in Ref.\cite{Dav1}.

Numerical results of the obtained equations are presented in terms of temperature, the DDI, and the gas parameter in the weak-coupling regime.
We show that the normal and anomalous fractions increase with $r_s$. 
In a comparison with a conventional dipolar BEC this study reveals that the DDI can decrease both the noncondensed and the anomalous concentrations
apart in the regime of very weak coupling where the depletion rises with the DDI.
Our results indicate also that the condensed fraction reduces with $r_s$ for any value of temperature.
Crucially, we point out  that the correction to the EoS arising from the quantum fluctuations exhibits an unconventional behavior with temperature, DDI and the gas parameter.
Furthermore, it is shown that the static structure factor overshoots unity displaying a sharp peak at lower temperatures and for relatively large DDI. 
It is found that the interplay of the Coulomb interaction and the DDI may lead to shift the height and the position of such peaks.
To the best of our knowledge, this is the first work unveiling these spectacular properties of CBG.

%Since that time there has been an explosion of experimental and theoretical interest worldwide in the study of ultradilute quantum droplets.

%based on dielectric theories \cite{Sing,Vash,Hir}, quantum-to-classical mappings \cite{Perr,Sand}. 
%Collective excitations and screening properties of CBG have been discussed in Refs. \cite{Alex1,Alex2}.
%Other methods have been used to describe the CBG  at $T=0$ based on dielectric theories \cite{Sing,Vash,Hore,Hir}  

%\section{ HFB theory for a charged Coulomb Bose gas with DDI}

We consider a gas of $N$ identical charged bosons with charge $e$ and mass $m$ in a box of volume $V$ with both contact and dipolar interactions moving
in a static uniform neutralizing background.
We assume that the dipoles are strictly aligned  along the $z$-axis, in this case the interaction potential has a contact component related to the $s$-wave scattering length $a$
and a dipolar component (see below).  
In the frame of the HFB theory, uniform charged bosons with DDI are described by the following nonlocal generalized  Gross-Piteavskii equation \cite{Dav,Boudj4,Boudj2,Boudj3}:
\begin{widetext}
	\begin{align} \label{NGPE}   
		i\hbar \dot{\Phi} ({\bf r},t) =  \bigg (-\frac{\hbar^2\nabla^2}{2m} -\mu\bigg) \Phi ({\bf r},t) +\int d{\bf r'} V({\bf r}-{\bf r'})  
		 \bigg [ n ({\bf r'},t) \Phi({\bf r},t) +\tilde n ({\bf r},{\bf r'},t)\Phi ({\bf r'},t)  +\tilde m({\bf r},{\bf r'},t)\Phi^*({\bf r'},t) \bigg ], 
	\end{align}
\end{widetext}
where $\Phi ({\bf r})=\langle \hat\psi ({\bf r})\rangle$ is the condensate wavefunction, with $\hat\psi ({\bf r})$ being the boson field operator, $\mu$ is the chemical potential, 
$n_c({\bf r})=|\Phi({\bf r})|^2$, $\tilde n ({\bf r})= \langle \hat {\bar\psi}^\dagger ({\bf r}) \hat {\bar\psi} ({\bf r}) \rangle $ and 
$\tilde m ({\bf r})= \langle \hat {\bar\psi} ({\bf r}) \hat {\bar\psi} ({\bf r}) \rangle $ are  respectively the condensed, noncondensed and anomalous densities, 
where $\hat{\bar \psi}({\bf r})=\hat\psi({\bf r})- \Phi({\bf r})$ is the noncondensed part of the field operator. 
The total density is given by $n({\bf r})=n_c({\bf r})+\tilde n ({\bf r})$. 
The terms $\tilde n ({\bf r, r'})$ and $\tilde m ({\bf r, r'})$ are respectively, the normal and the anomalous one-body density matrices
which account for the exchange interaction between the condensed and noncondensed atoms.
The two-body interaction potential is 
\begin{align}  \label{IntraPot}
	V ({\bf r}-{\bf r'})&= V_g ({\bf r}-{\bf r'})+V_{\text{dd}} ({\bf r}-{\bf r'})+V_c ({\bf r}-{\bf r'}) \\
&=g\delta({\bf r}-{\bf r'})+\frac{C_{dd}} {4\pi} \frac{1-3\cos^2 \theta} {|{\bf r}-{\bf r'}|^3}+ \frac{C_c} {|{\bf r}-{\bf r'}|},\nonumber
\end{align}
where $g=4\pi \hbar^2 a/m>0$ is the coupling constant corresponds to the contact interaction with $a$ being the $s$-wave scattering length, 
$C_{dd}=d^2/\epsilon_0$ is the electric DDI strength with $\epsilon_0 $ being the permittivity of vacuum,
$ \theta$ is the angle between the polarization direction and $\bf r$,
$C_c$ describes the strength of  Coulomb interactions, it is related to the Bohr radius via: $C_c=4\pi \hbar^2/(m a_B)$ \cite{Foldy,Dav1,Alex1}.
It is clear that for $C_c=C_{\text{dd}}=0$,  Eq.(\ref{NGPE}) reduces to the standard local Gross-Piteavskii equation.

Now we calculate the elementary excitations and fluctuations of a CBG.
In the high-density limit and when the temperature is close to zero, we can linearize Eq.(\ref{NGPE}) using the
$\Phi = \sqrt{n_{c}}+\delta \Phi $, where $\delta \Phi ({\bf r},t)= u_{k}  e^{i {\bf k \cdot r}-i\varepsilon_k t/\hbar}+v_{k} e^{i {\bf k \cdot r}+i\varepsilon_k t/\hbar} \ll \sqrt{n_{c}}$,
where $u_k$ and $v_k$ are the Bogoliubov amplitudes. 
In Fourier space, the wavefunctions are real-valued  ($\Phi_{0}=\Phi_{0}^*=\sqrt{n_{c}}$),  and the interaction potential (\ref {IntraPot}) is written as:
\begin{align}  \label{IntraPotFour}
\tilde V (\mathbf k)&=\tilde V_g (\mathbf k)+\tilde V_{\text{dd}} (\mathbf k)+\tilde V_c (\mathbf k) \\
&=g \big[1+\epsilon_{\text{dd}} (3\cos^2\theta_k-1)+\epsilon_c/k^2\big], \nonumber
\end{align}
where $\epsilon_{\text{dd}}=C_{\text{dd}}/3g$ is the relative strength which describes the interplay of contact interaction and the DDI, 
$\theta_k$ is the angle between the vector $\mathbf k$ and the polarization direction, and 
$\epsilon_c=C_c/g$ which has dimension (length)$^{-2}$ is the relative coupling strength which describes the interplay of contact interaction and Coulomb interaction.
For the most widely utilized species of cold atoms (such as Rb, Cr, Er, Dy), $\epsilon_c a_B^2$ ranges from 0.005 to 0.01.
Due to  the electroneutrality, one can set $\tilde V_c (\mathbf k \equiv 0)=0$, which is a consequence of the compensation of the boson-boson repulsion by the
attraction due to a spatially homogeneous charged background \cite{Alex2}.

The chemical potential is given according to Eq.(\ref{NGPE}) by \cite{Boudj5,Yuk} 
\begin{align} \label{chim0}
\mu=\tilde V(|\mathbf k|=0) n + \frac{1}{V}\sum\limits_{\bf k \neq 0} \tilde V(\mathbf k) \big(\tilde n_k +\tilde m_k \big),
\end{align}
where $\tilde n_k$ and  $\tilde m_k$  stand for the normal and anomalous distributions which can be defined in the spirit of the HFB approximation as \cite{Dav,Yuk,Boudj5}:
\begin{equation}\label{Ndiso} 
\tilde n_k=[v_k^2+ (u_k^2+v_k^2)N_k],
\end{equation}
and 
\begin{equation}\label{Mdiso} 
 \tilde m_k=u_k v_k (2N_k+1),
\end{equation}
where $N_k=\langle \hat b_k^{\dagger} \hat b_k\rangle=[\exp(\varepsilon_k/T)-1]^{-1}$  are occupation numbers for the excitations. 
The solution of the resulting Bogoliubov-de-Gennes (BdG) equations gives for the Bogoliubov quasiparticle amplitudes
$$
u_k^2 = \frac{\omega_k+\varepsilon_k}{2\varepsilon_k}, \qquad
v_k^2 = \frac{\omega_k-\varepsilon_k}{2\varepsilon_k},
$$
and for the Bogoliubov excitations energy \cite{Boudj5,Yuk} 
\begin{equation} \label{Bog11}
\varepsilon_k =\sqrt{\omega_k^2 -\Delta_k^2},
\end{equation}
where 
$$
\omega_k \equiv E_k + n  \tilde V (|\mathbf k|=0) + n_c  \tilde V ({\bf k}) +\frac{1}{V} \sum_{p\neq 0} \tilde n_p  \tilde V ({\bf k}+{\bf p}) - \mu_1,
$$
and
$$
\Delta_k \equiv n_c  \tilde V ({\bf k}) +\frac{1}{V} \sum_{p\neq 0} \tilde m_p  \tilde V ({\bf k}+{\bf p}),
$$
where $ E_k= \hbar^2 k^2/ 2m$ is the free particle energy.\\
Evidently, the HFB spectrum (\ref{Bog11})  has an unphysical  gap in the limit of long wavelengths due to the inclusion of the anomalous correlations.
To circumvent this problem, we define the chemical potential $\mu_1$ as \cite {Griffin,Yuk, Boudj}:
\begin{equation}
\label{Chim1}
\mu_1= n  \tilde V(|\mathbf k|=0)+ \frac{1}{V}  \sum_{k\neq0} \tilde V ({\bf k}) (\tilde n_k -\tilde m_k).
\end{equation}
with the condition of charge neutrality which cancels the $k= 0$ term associated with Coulomb interaction in the sum \cite{Dav,Alex2}.
It is obvious that the chemical potential $\mu_1$ of Eq.(\ref{Chim1}) renders the spectrum (\ref{Bog11}) gapless in agreement with
the Hugenholtz-Pines theorem \cite{HP}. Importantly,  the excitation spectrum (\ref{Bog11}) has a roton-maxon structure originating from the 
anisotropy of the Hartree-Fock corrections.

\begin{figure}
	\centering 
	\includegraphics[scale=0.75, angle=0] {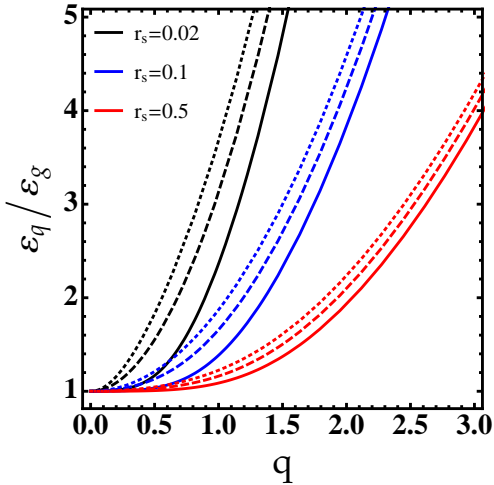}
	\caption { Bogoliubov spectrum for $\theta=\pi/2$ and for different values of $r_s$ and $\epsilon_{dd}$ at $r_0/\xi=\sqrt{0.5}$ or equivalently $na^3=0.0011$.
Solid lines: $\epsilon_{dd}=0.95$. Dashed lines: $\epsilon_{dd}=0.45$.  Dotted lines: $\epsilon_{dd}=0$.}
	\label{BogS}
\end{figure}

Setting $g=C_{\text{dd}}=0$ in Eq.(\ref{Bog11}), the HFB excitation energy for a charged Bose gas is well reproduced \cite{Dav}.
In the long-wavelength limit $k \rightarrow 0$ where $\lim\limits_{k\rightarrow 0} (\tilde n_k -\tilde m_k) \simeq -1/2$ \cite{Dav}, 
the zero-temperature Bogoliubov excitations energy (\ref{Bog11}) coincides with the plasma energy  $\varepsilon_g/\hbar= \sqrt{nC_c/m}$ (i.e. plasmon gap).
At finite temperatures, one can expect that the spectrum energy (\ref{Bog11}) increases with $T$ and vanishes at the transition.
In the high momenta limit, $k\rightarrow \infty$, the excitations spectrum (\ref{Bog11}) reduces to the free particle law ($\varepsilon_k=E_k$).

For the sake of simplicity, we shall assume from now on that $\tilde{m}/n_c \ll 1$ and $\tilde{n}/n_c \ll1$ (i.e. we neglect higher-order terms)  and set $\theta_k = \pi/2$. 
In such a case the Bogoliubov excitations energy (\ref{Bog11}) reduces to the following dimensionless form:
\begin{equation}
\varepsilon_{q}=\varepsilon_g \sqrt{ \frac{q^4} {12r_s} +\frac{q^2} {3r_s} \frac{r_0^2} {\xi^2}(1-\epsilon_{dd})  +1 },
\end{equation}
where $q=k r_0$, and  $\xi= \hbar/\sqrt{mng}$ is the standard healing length of the condensate.

In Figure \ref{BogS} we show the behavior of the  Bogoliubov spectrum for different values of the coupling parameter $r_s$.
It is clearly seen that the spectrum $\varepsilon_{q}$ increases monotonically with $q$ whatever the values of $r_s$ and $\epsilon_{dd}$.
It increases with decreasing both $r_s$ and $\epsilon_{dd}$.
%One can expect that the effect of the dipole-dipole orientation on the excitations spectrum could be important giving rise to affect the and the thermodynamics and the coherence of the system.

%\section{Fluctuations and equation of state}

In a homogeneous Bose gas, the depletion  and the anomalous density  are defined as \cite{Dav,Boudj5,Yuk}:
$\tilde{n}=V^{-1} \sum\limits_{\bf k \neq 0} n_k$, and $\tilde{m}=-V^{-1} \sum\limits_{\bf k \neq 0} \tilde m_k$.
Working in the thermodynamic limit, the sum over $k$ can be replaced  by integrals according to the
prescription $\sum\limits_{\bf k} \rightarrow V\int_{ 0}^{ \infty}  d{\bf k}/(2\pi)^3 $. 
Thereafter inserting the identity $2N (x)+1= \coth (x/2)$ into Eqs.(\ref{Ndiso}) and (\ref{Mdiso}), we then obtain for the noncondensed and anomalous densities \cite{Boudj1,Boudj5}:
\begin{align}\label{Dep} 
\tilde n=\frac{1}{2}\int \frac{d {\bf k}}{ (2\pi)^3} \left[\frac{E_k+n \tilde V(\mathbf k)} {\varepsilon_k}\text {coth}\left(\varepsilon_{k}/2T\right)-1\right],
\end{align}
and
\begin{equation}\label{mDep} 
\tilde m =-\frac{1}{2}\int \frac{d {\bf k}}{ (2\pi )^3} \frac{n \tilde V(\mathbf k) } {\varepsilon_k}\text {coth}\left(\varepsilon_{k}/2T\right).
\end{equation}
The validity of the present HFB approach requires the inequality: $\tilde n \ll n$. 
This implies that to have a dilute CBG, the following conditions: $r_s \ll 1$ and $na^3\ll1$ must be fulfilled. \\
The condensed fraction can be evaluated through: $n_c/n=1-\tilde n/n$. 
For $g=C_{dd}=0$, the condensed fraction reduces to $n_c/n \simeq 1-0.2\, r_s^{3/4}$ \cite{Foldy}.

Correction to the chemical potential due to the Lee-Huang-Yang (LHY) quantum fluctuations can be given from Eq.(\ref{chim0}) as  \cite{Boudj1,Boudj5}:
$\mu_{\text{LHY}}= V^{-1}\sum\limits_{\bf k \neq 0} \tilde V(\mathbf k) \big(\tilde n_k +\tilde m_k \big)$.
Then using the definitions $\tilde n_k$ and  $\tilde m_k$ from Eqs.(\ref{Ndiso}) and (\ref{Mdiso}), we obtain:
\begin{align} \label{chim01}
\mu_{\text{LHY}}=\frac{1}{2}\int \frac{d {\bf k}}{ (2\pi)^3} \tilde V(\mathbf k) \left[\frac{E_k} {\varepsilon_{k}}\text {coth}\left(\varepsilon_{k}/2T\right)-1\right]. 
\end{align}
This equation permits us to calculate the LHY corrections to all thermodynamic quantities.
In the absence of the short-range and dipolar interactions the LHY-corrected EoS reads $\mu_{\text{LHY}} \simeq-0.23\, \varepsilon_g r_s^{3/4}$ \cite{Foldy}.

\begin{figure}
	\centering 
	\includegraphics[scale=0.72, angle=0] {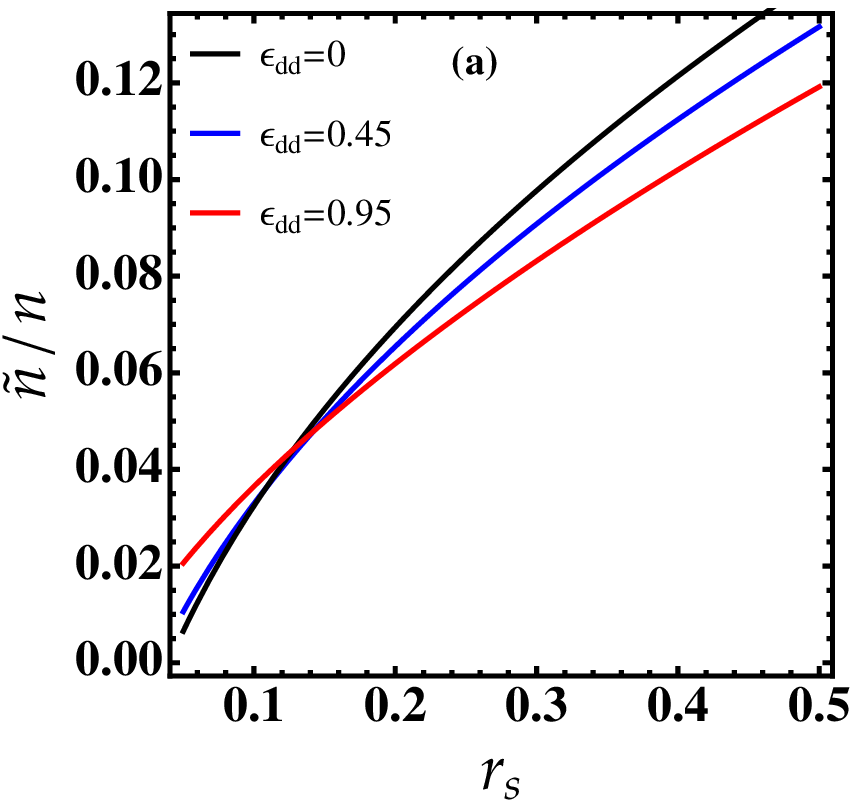}
	\includegraphics[scale=0.72, angle=0] {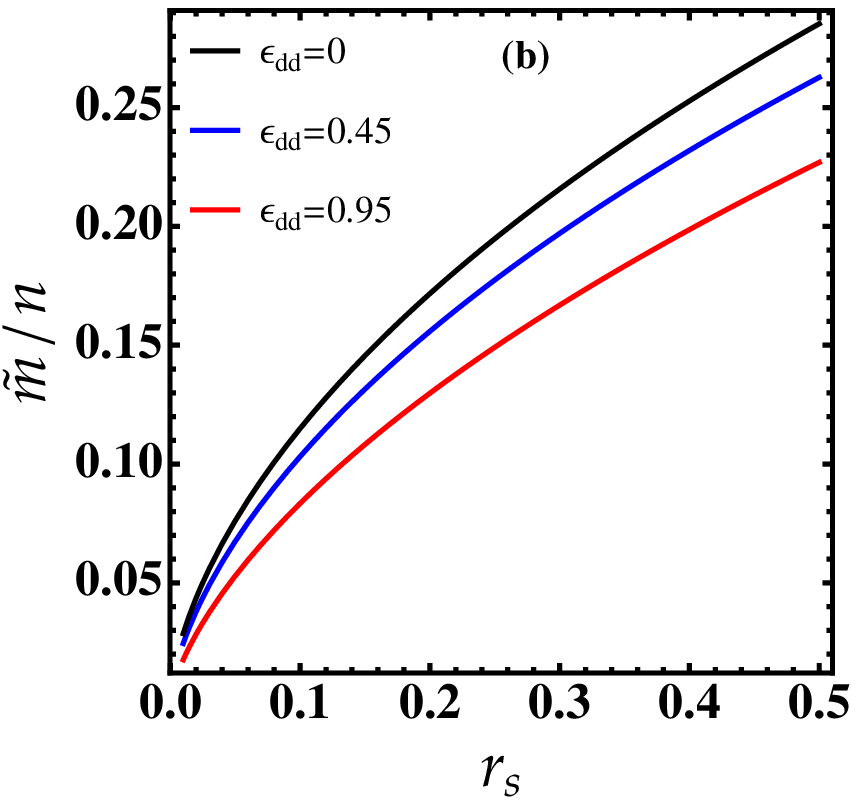}
	\caption { (a) Depletion $\tilde n/n$ at $T=0$ as a function of the coupling parameter $r_s$ for different values of $\epsilon_{dd}$.
(b) Anomalous fraction $\tilde m/n$ at $T=0$ as a function of $r_s$ for different values of $\epsilon_{dd}$. Here we set $na^3=0.0011$.}
	\label{Fl}
\end{figure}

It is worth stressing that the exact analytical solutions of the integrals (\ref{Dep}), (\ref{mDep}) and (\ref{chim01}) are not trivial except in some limiting cases. 
Therefore, we solve them numerically. 

The results for the noncondensed and the anomalous fractions for different values of $r_s$ and $\epsilon_{dd}$ 
are shown graphically in Fig.\ref{Fl}. 
We see that the noncondensed and the anomalous fractions increase with the coupling strength $r_s$ even in the absence of the DDI ($\epsilon_{dd}=0$).
In the very weak coupling regime $r_s \lesssim 0.15$,  $\tilde n/n$ rises with $\epsilon_{dd}$ (see the inset of Fig.\ref{Fl}.a).
On the contrary, for $r_s  > 0.15$, the depletion decreases with the DDI even for relatively large $\epsilon_{dd}$.
This can be attributed to the effect of the Coulomb interaction which dominates both the short-range and the dipolar interactions.
The anomalous fraction is decreasing with the DDI in the whole range of $r_s$ as is seen in Fig.\ref{Fl}.b.
For very small $r_s$, one has $\tilde n/n \ll 1$ and $\tilde m/n \ll 1$  in good agreement  with the results of the Bogoliubov \cite{Dav}
and with Monte Carlo data \cite{Mor}. 
Whereas for larger $r_s$, both the depletion and the anomalous fraction increase continuously whatever the value of $\epsilon_{dd}$. 
For example, for  $r_s =0.5$ and $\epsilon_{dd}=0.45$, one has $\tilde n/n >  10\%$ and $\tilde m/n \gtrsim 25\%$,
pointing out that the present HFB theory becomes no longer applicable in the regime of $r_s \gtrsim 0.5$.
Another important remark is that the anomalous fraction is larger than the normal fraction similarly to the neutral atomic dipolar and nondipolar BECs \cite{Yuk, Boudj5,Boudj,Griffin,Boudj1}. 

\begin{figure}
	\centering 
	\includegraphics[scale=0.7, angle=0] {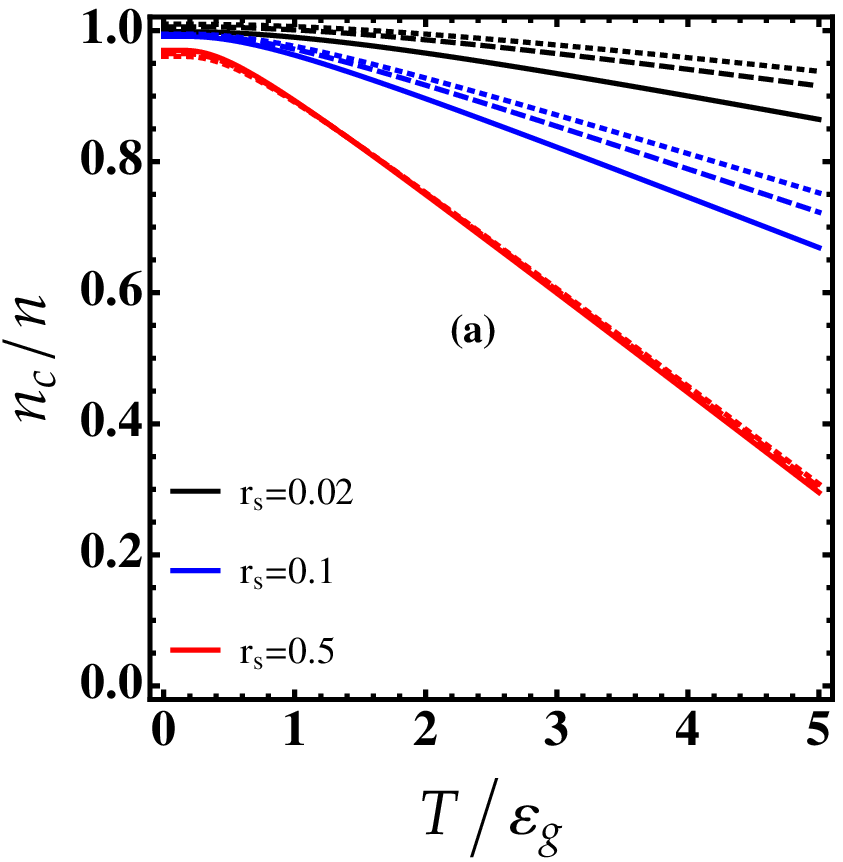}
	\includegraphics[scale=0.7, angle=0] {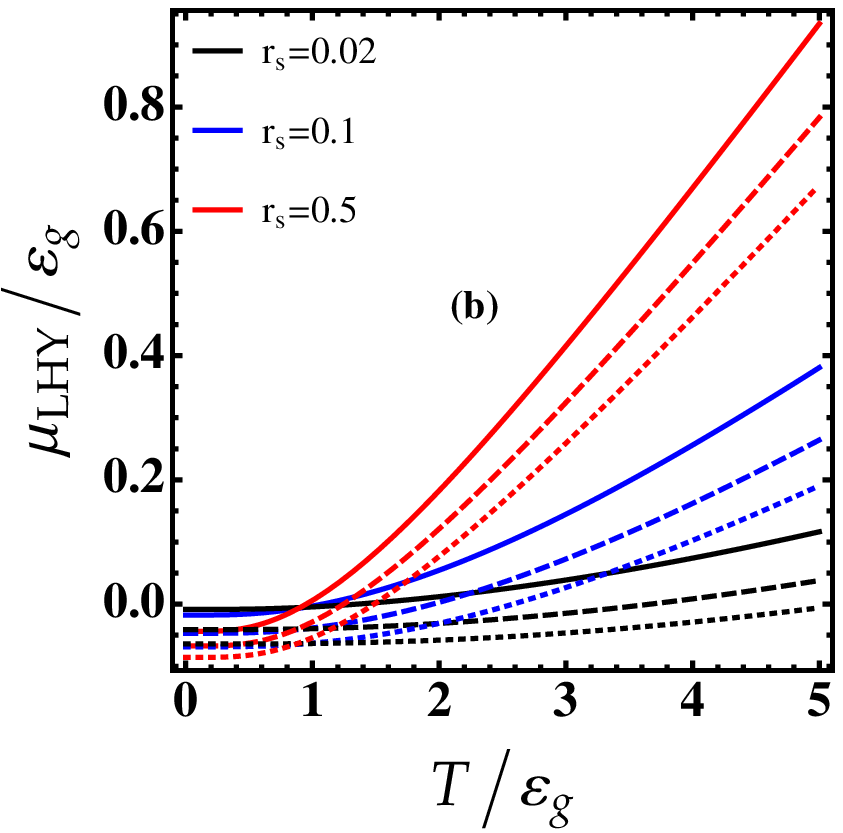}
	\caption { (a) Condensate fraction $n_c/n$ as a function the temperature $T/\varepsilon_g$ for different values of $r_s$ and $\epsilon_{dd}$.
(b) LHY-corrected EoS  $\mu_{\text{LHY}}/\varepsilon_g$ as a function the temperature $T/\varepsilon_g$ for different values of $r_s$ and $\epsilon_{dd}$.
Solid lines: $\epsilon_{dd}=0.95$. Dashed lines: $\epsilon_{dd}=0.45$. Dotted lines: $\epsilon_{dd}=0$. Here we set $na^3=0.0011$.}
	\label{FS}
\end{figure}

Figure \ref{FS}.a reports the condensate fraction $n_c/n=1- \tilde n/n$ as a function of temperature $T/\varepsilon_g$  for different values of the coupling strength $r_s$ 
and of the relative interaction strength $\epsilon_{dd}$. 
It is clearly visible that the condensate fraction strongly decreases with $r_s$ for any values of temperature.
This indicates that the dipolar CBG becomes strongly depleted for large $r_s$ and $\epsilon_{dd}$ (see red lines).
We see also that at fixed temperature,  $n_c/n$ lowers with increasing $\epsilon_{dd}$ except for large values of $r_s$, where it slightly decreases with decreasing $\epsilon_{dd}$
due to the interplay of the Coulomb and dipolar interactions.

Figure \ref{FS}.b depicts that at low temperatures $T \lesssim \varepsilon_g$, the LHY-corrected chemical potential reduces with $r_s$ and $\epsilon_{dd}$.
Remarkably, it remains negative in such a regime leading to decrease the total EoS. The negative value of $\mu_{\text{LHY}}$ is a product of oppositely charged background \cite{Alex1}.
At  $T \gtrsim \varepsilon_g$, $\mu_{\text{LHY}}/\varepsilon_g$ increases linearly with temperature especially for large $r_s$ regardless of the value of  $\epsilon_{dd}$.
This can be attributed to the competition between the contact, the dipole-dipole and Coulomb interactions.

%\section{Static structure factor}\label{SSF}

\begin{figure}
	\centering 
	\includegraphics[scale=0.7, angle=0] {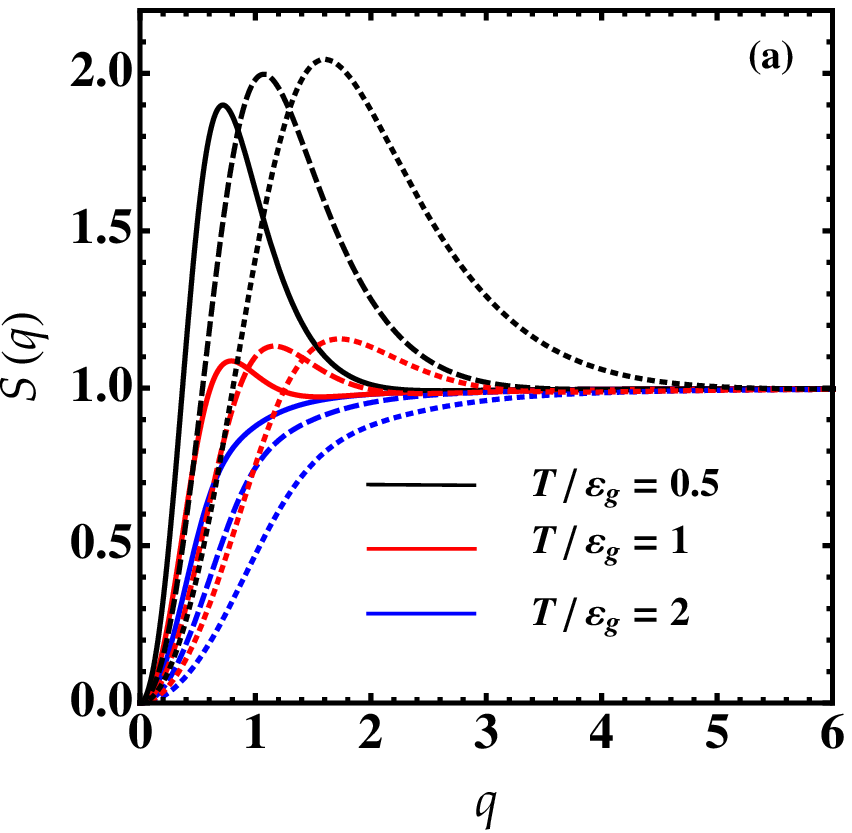}
	\includegraphics[scale=0.7, angle=0] {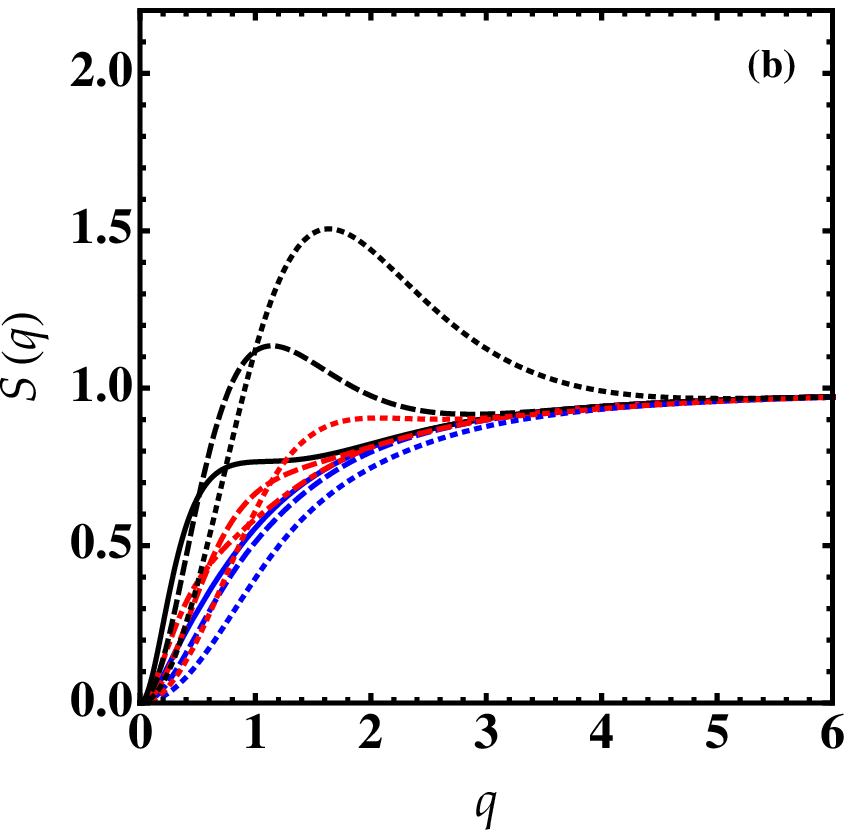}
	\caption { Static structure factor from Eq.(\ref{SFac}) as a function of dimensionless variable $q=kr_0$ for different values of  $T/\varepsilon_g$ and $r_s$.
Parameters are: $na^3=0.0011$, $\epsilon_{dd}=0.95$ (a) and  $\epsilon_{dd}=0$ (b).
Solid lines: $r_s=0.02$. Dashed lines: $r_s=0.1$.  Dotted lines: $r_s=0.5$. }
	\label{SF}
\end{figure}

Information on the coherence and on the fluctuations of CBG are contained in the static structure factor which is defined as 
the Fourier transform of the density-density correlation function $S({\bf k})= \langle \delta \hat n({\bf k}) \delta \hat  n(-{\bf k})\rangle/n$ \cite{PitaevString},
where $\delta \hat n ({\bf k})=\int d{\bf r} \,\delta \hat n ({\bf r}) e^{-i.\bf k.r}$ and 
$\delta \hat n({\bf r})=\sqrt{n ({\bf r})} \sum_k \big\{[u_k({\bf r})-v_k ({\bf r})] \exp (-i \varepsilon_k t/\hbar ) \, \hat {b}_{k}+  H.c\big\}$.
In terms of the elementary excitation energy $S({\bf k})$ reads \cite{PitaevString}:
\begin{equation}\label {SFac}
S({\bf k})= \frac{E_k}{\varepsilon_{ k}} \coth \left(\frac{\varepsilon_{ k}}{2T}\right), 
\end{equation}  
At zero temperature, Eq.(\ref{SFac}) reduces to $ S({\bf k})= E_k/\varepsilon_{k}$.
At low temperatures and in the phonon regime ($k\rightarrow 0$) one has $S({\bf k}) \simeq T/\varepsilon_g$.
In the opposite situation, at higher $T$,  $S({\bf k})$ simplifies to its zero temperature value except in the limit $k\rightarrow 0$ where 
the structure factor approaches its asymptotic value \cite{PitaevString}.
A non-correlated gas has a structureless spectrum $S(k)=1$. 

The numerical solution of Eq.(\ref{SFac}) is presented in Fig.\ref{SF}. 
As one can see from the figure, at low temperatures $T< \varepsilon_g$ where the main contribution comes from low momenta,  
the static structure factor has a strong dependence on the temperature, the Coulomb interaction and on the DDI.
We observe also that S({\bf k}) increases significantly and develops a pronounced peak around $q=q_0$.
For instance, for $r_s=0.5$, $q_0 \simeq 1.6$ or equivalently $k \simeq 1.6/r_0$ (see Fig.\ref{SF}.a). The position of such a peak relies on $r_s$. 
Remarkably, S({\bf k}) develops a  peak even at relatively high temperatures ($T \simeq \varepsilon_g$) for all $r_s$ due to the interplay of Coulomb and dipolar interactions.
Augmenting further the temperature  ($T \geq 2 \varepsilon_g$), thermal effects do not favor such a localization behavior of the particles 
and the static structure factor is almost monotonic (see blue lines in Fig.\ref{SF}).

However in the case of nondipolar CBG ($\epsilon_{dd}=0$), the  structure factor becomes less significant and exhibits a peak only at low temperatures and
for $r_s > 0.02 $ in contrast to the dipolar CBG (see the Fig.\ref{SF}.b).
This can be understood due to the fact that in the absence of DDI and for $r_s > 0.02$, the Coulomb interaction dominates the system leading 
to strong thermal fluctuations even at $T \lesssim \varepsilon_g/2$, giving rise to destroy the coherence of the system.

%\section{Conclusion}\label{conc}

In conclusion, we studied the ground-state properties of a weakly interacting CBG with DDI at both zero and finite temperatures using the
self-consistent HFB theory. We derived the generalized nonlocal Gross-Pitaevskii equation
that describe the dynamics and the equilibrium of such a system.
By solving the BdG equations we analyzed the Bogoliubov excitations spectrum that exhibits a plasmon gap giving rise to infrared divergences arising from the Coulomb interaction. 
Furthermore, the normal and anomalous fractions, the EoS, and the static structure factor have been computed numerically. 
We show that the intriguing interplay of Coulomb interactions and the DDI leads to affect these quantities 
and thus plays a pivotal role in the physics of the system.

Even though the experimental realization of identical charged bosons with DDI is a challenging problem, 
they constitute a promising area for applications. 
In the limit of a strong (even intermediate) coupling, one can expect that the presence of the DDI may lead to the appearance of a very sharp peak in the static structure factor
driving the system to a transition to a Wigner crystal phase.  A qualitative study of this crystallization necessitates  sophisticated tools such as Quantum Monte Carlo simulations.
An important extension of our work would be the study of the Meissner-Ochsenfeld effect in dipolar CBG.
The competition between repulsive short-range, Coulomb and dipolar interactions may enhance the coherence of the system
implying a singularity in the susceptibility prior to the BEC phase \cite{Shun}. 
This could be a signature of the emergence of such a Meissner-Ochsenfeld effect.

%\section{Acknowledgements}

%\subsection*{Data availability statement}
%The data generated and/or analyzed during the current study are not publicly available for legal/ethical reasons
%but are available from the corresponding author on reasonable request.

\end{document}